# Length Scale Dependence of Periodic Textures for Photoabsorption Enhancement in Ultra-thin Silicon Foils and Thick Wafers

K Kumar[1], A Khalatpour[2], G Liu[1], J Nogami[1*] and N P Kherani[1, 2]

[1] Department of Materials Science and Engineering, University of Toronto, Toronto, M5S 3E4, Canada

[2] Department of Electrical and Computer Engineering, University of Toronto, Toronto, M5S 3G4, Canada

*jun.nogami@utoronto.ca

**Abstract**

In this paper, we simulate a front surface inverted pyramidal grating texture on 2 to 400 μm thick silicon and optimize it to derive maximum photocurrent density from the cell. We identify a "one size fits all" front grating period of 1000 nm that leads to maximum photo-absorption of normally incident AM1.5g solar spectrum in silicon (configured with a back surface reflector) irrespective of the thickness of the crystalline silicon absorbing layer. With the identification of such universally optimized periodicity for the case of an inverted pyramidal grating texture, a common fabrication process can be designed to manufacture high-efficiency devices on crystalline silicon regardless of wafer thickness. In order to validate the results of the simulation, we fabricated high resolution inverted pyramidal textures on a 400 μm thick silicon wafer with electron beam lithography to compare the reflectance from submicron and wavelength scale periodic textures. The experimental reflectance measurements on textures confirm that a 1000 nm period grating texture performs better than a 500 nm period texture in reducing reflectance, in agreement with the simulations.

1. Introduction

In order to meet grid-parity and to compete with energy derived from fossil fuels, the crystalline silicon (c-Si) photovoltaic (PV) industry faces the challenge of lowering the cost of installed PV panels while retaining their high solar-to-electrical energy conversion efficiencies. One avenue of reducing the cost of c-Si based PV panels is to reduce the thickness of their constituent cells; the material cost of c-Si wafers represents ~ 35 percent of the cost of the installed panel[1-3]. In this context, a c-Si wafer which is typically 180 μm in thickness[1, 2, 4] could be replaced with less expensive thinner c-Si foils of thickness < 20 μm[5]. However, increased optical losses in PV cells made from thin c-Si foils necessitates light trapping schemes to be integrated into their design in order to achieve high efficiencies in addition to minimizing Fresnel reflections at the front air-silicon interface.

In the geometric optics regime, the most effective approach to trap light in thin Si wafers is to incorporate Lambertian scattering at the front or both front and back surfaces and an ideal reflector at the back surface. For this type of structure, geometrical optics theory predicts that the optical path length near the absorption band edge (i.e., in the weakly absorbing region) can be increased by a maximum factor of $4n^2$ as compared to a plain untextured wafer, where $n$ is the refractive index of the absorbing medium. This $4n^2$ enhancement limit, also known as the Lambertian or Yablonovitch limit[6-8], suggests the possibility of a ~50 fold increase in the path length in silicon ($n \sim 3.5$) and if achieved, the prospect of a few micron thick c-Si solar cell with similar efficiencies to current 180 μm thick c-Si PV cells.

A periodically structured surface or grating is one type of optical structure that can be used to enhance light absorption within a bare silicon slab. The application of such texture was first proposed by Sheng *et al.* in 1983 for enhancing the absorption of light in amorphous-silicon thin film cells[9] and later studied by Heine and Morf for wafer-based high efficiency c-Si cells[10]. Since then, grating structures with hemispherical[11], triangular[12], conical[13], pyramidal[14, 15], inverted pyramidal[16], nano-holes[17] and rod-like[17] features have been studied, with the objective of maximizing optical absorption within a cell. Of particular relevance to this work are prior studies which have shown that a high performance grating structure consists of tapered features highly packed in a two dimensional array with grating period slightly smaller than the wavelength range of interest[11, 15, 18]. The tapered features reduce reflection losses at the front surface by producing a refractive index, $n$, profile that gradually increases from $n = 1$ in air to $n = 3.5$ in the absorber (silicon). Furthermore, the periodic profile of the index of refraction in the lateral direction couples light into diffractive modes that propagate at large off-normal angles within the silicon wafer, thereby increasing the optical path-length through the cell. Light propagation in these diffractive modes cannot be described by geometric optics and in this case coherent wave optics analysis is required. In the wave optics regime the Lambertian limit is no longer strictly applicable and absorption can surpass the Lambertian limit in a limited spectral range at normal incidence at the cost of absorption at other incident angles, keeping the angle-averaged absorption below the Lambertian limit [9, 15, 18, 19].

Another tapered surface texture that has been studied in the literature is an inverted pyramidal grating, which also offers good light trapping and minimally increases the surface area upon texturing which is desirable for achieving low surface recombination losses. The inverted pyramidal features also exhibit graded refractive index profile from air to silicon which serves to decrease reflection losses. Furthermore, since the fabrication steps are easy to implement, inverted-pyramidal grating textures have been applied on both thick and thin crystalline silicon wafers using various fabrication methods[16, 20-23] to demonstrate high efficiency PV cells [16, 20, 24]. Prior theoretical and experimental studies on inverted pyramid grating texture have focused on specific grating periodicities and feature size which are either very thick (~ 400 μm)[22, 23, 25] or very thin silicon membranes (< 50 μm)[16, 24]. For example, in 2009, an energy conversion efficiency of 23% was calculated by Kray *et al.* for a 40 μm thick high quality c-Si solar cell with an inverted pyramid grating of period $\Lambda \sim 2$ μm[24] and recently Mavrokefalos *et al.* employed a 700 nm period inverted pyramidal grating to produce ~25% efficient cells on 10 μm thick c-Si foils[18]. However, heretofore the understanding of the precise interplay between the grating periodicity, wafer thickness, and absorption enhancement in c-Si wafers with an inverted pyramidal texture is still unclear.

2. **Design and Optimization**

We use a wave-optical approach to study inverted pyramidal textures with grating periods ranging from subwavelength to ~ 2 × typical wavelengths (300 – 1100 nm) on the front surface of 2 μm to 400 μm thick crystalline silicon with the objective of maximizing photoabsorption of the AM1.5g solar spectrum. We first study absorption in a planar silicon slab of thickness ranging from 2 – 400 μm without and with a perfect back reflector (a perfect mirror) under front-side illumination, as shown in Figure 1(a) and (b), respectively, to identify the weakly absorbing spectral region. This sets the reference point for further calculations. We then add an inverted pyramidal texture of grating period ($\Lambda$) at the front surface of the wafer and optimize the periodicity to reduce reflection and increase light trapping at the same time. In this case, we consider texture with fill fraction = 1, that is, where the inverted pyramids are abutting and there is no flat region (mesa) between the features, as shown in figure 1(c). We evaluate the performance of the grating texture based on the maximum calculated photocurrent density produced. The aspect ratio (width of inverted pyramid to its depth) is kept constant at 1.404, which is typical of inverted pyramids produced by anisotropic chemical etching of Si(001) wafers. The photoabsorption in textured silicon subjected to the normally incident AM1.5g solar spectrum[26] is calculated in the wavelength ($\lambda$) range from 280 - 1107 nm using the scattering matrix method[27, 28] (see supplementary information section for details) and the optical constants of silicon[29]. The photocurrent density is calculated with the assumption of 100% internal quantum efficiency, that is, it is assumed that each absorbed photon with energy greater than the silicon band gap produces one electron-hole pair. To ensure accuracy of the solution, the individual layer thicknesses and the number of modes in the calculations were set to $\lambda/20$ and 121, respectively. Subsequently we investigate photocurrent loss due to flat regions (mesas) in the texture with fill fraction < 1 (figure 1(d)), which are unavoidable for practical reasons, and optimize an anti-reflective coating on the front surface to overcome the optical losses due to the mesas (figure 1 (e)). Our work identifies a "one size fits all" front texture that leads to maximum photo-absorption in thin and thick wafers. With the identification of such texture, a common fabrication process can be designed to manufacture high-efficiency devices on c-Si wafers of various thicknesses in industry.

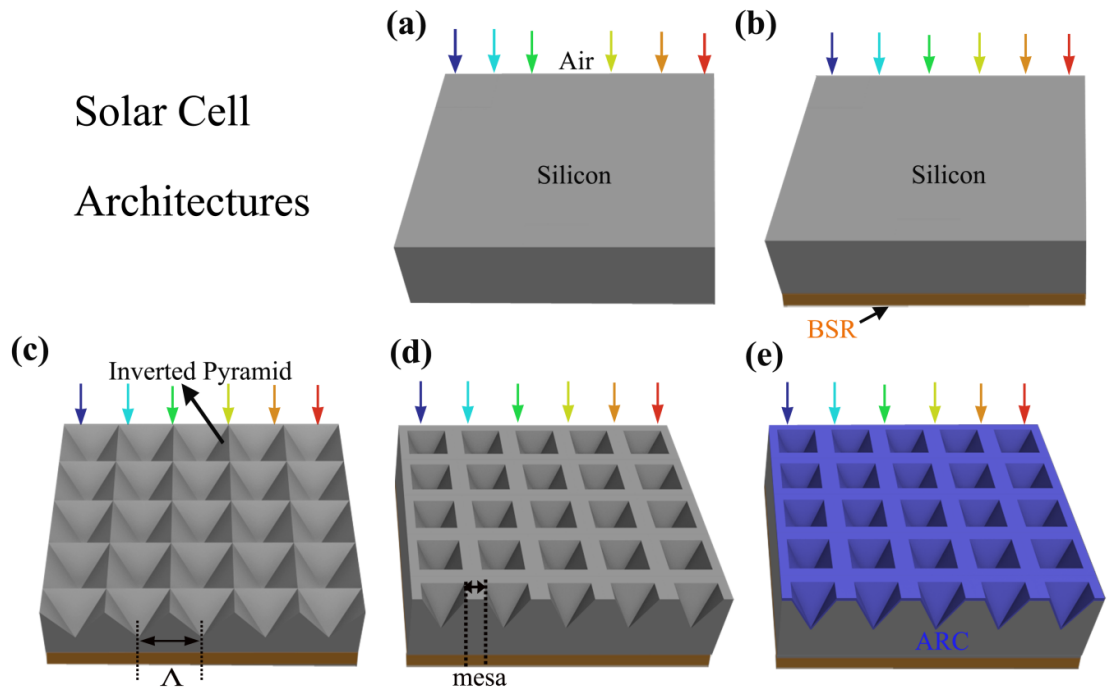

**Figure 1** Schematic depiction of the solar cell architectures investigated in this work. A planar silicon slab in air (a) without and (b) with a perfect back surface reflector (BSR, orange), with front inverted pyramid texture (c) without mesa (filling fraction = 1), (d) with mesa (fill fraction <1), and (e) with mesa and an antireflective coating (ARC, blue) and BSR. All architectures are subject to AM1.5g front side illumination.

First, we study spectral losses in a standard 400 μm thick silicon wafer to understand the optical loss mechanisms. Then we extend our knowledge to identify weakly absorbing spectral regions in thinner silicon wafers. This is particularly helpful in understanding the physics underlying the absorption enhancement by a grating texture. The spectral reflectance ($R$) and transmittance ($T$) for a 400 μm thick planar slab of crystalline silicon in 300 nm - 1107 nm wavelength region is shown in

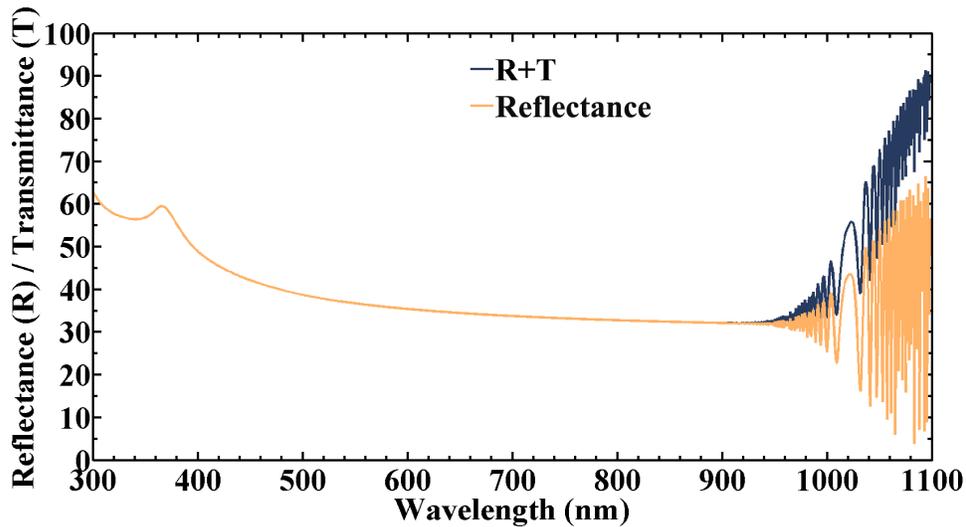

Figure 2.

Light with wavelength between 300 - 924 nm is almost completely absorbed ($T < 0.1\%$) and the optical losses of 39% over this spectral region is due to the reflection of light at the front air-silicon interface. The transmittance increases to ~ 16 % in the 924 - 1107 nm wavelength region, showing insufficient silicon thickness to absorb these wavelengths due to their large penetration depths. Such weakly absorbed light is partially reflected inside the slab at the rear silicon-air interface and partially transmitted, thus contributing to the optical losses in the slab. In this spectral region, the reflectance and transmittance plots in Figure 2 displays a series of peaks that arise due to Fabry-Perot interference patterns generated from the internal reflection of electromagnetic waves from the two interfaces bounding the silicon wafer. The presence of these Fabry-Perot resonances indicate the incomplete absorption of photons after a single pass through the Si wafer and in this case both transmission at the rear side of the cell as well as reflection contribute to the overall optical losses in the wafer. Thus, the front grating of a silicon solar cell needs to be designed to facilitate diffraction in weakly absorbing spectral regions where Fabry-Perot interference patterns appear in the reflection and transmission spectra. In contrast, the reflection spectra do not exhibit Fabry-Perot interference patterns for wavelengths between 300 – 924nm, which is indicative of the fact that the optical path length through the 400 μm thick Si wafer is sufficient to strongly absorb photons on their first pass through the wafer. However, it is noted that the graded index profile of the front texture enhances absorption over this spectral region by reducing reflection losses.

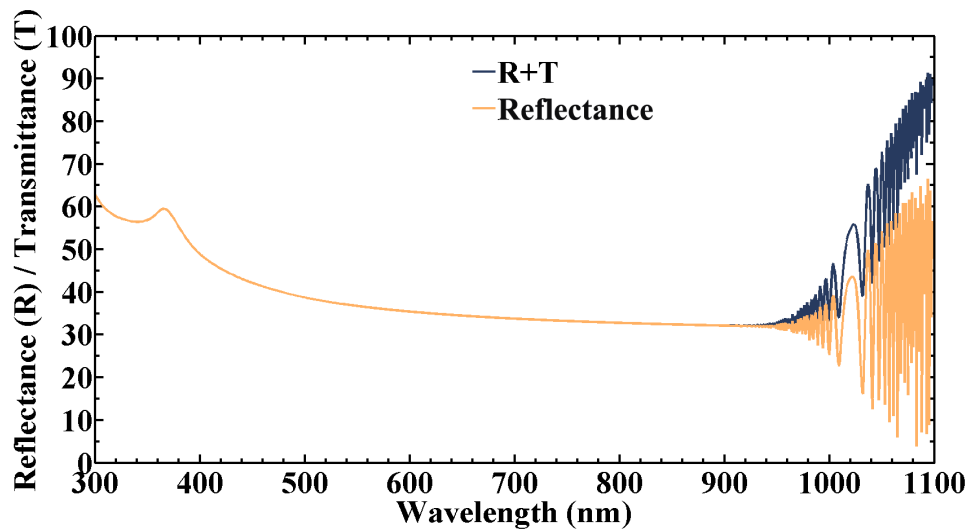

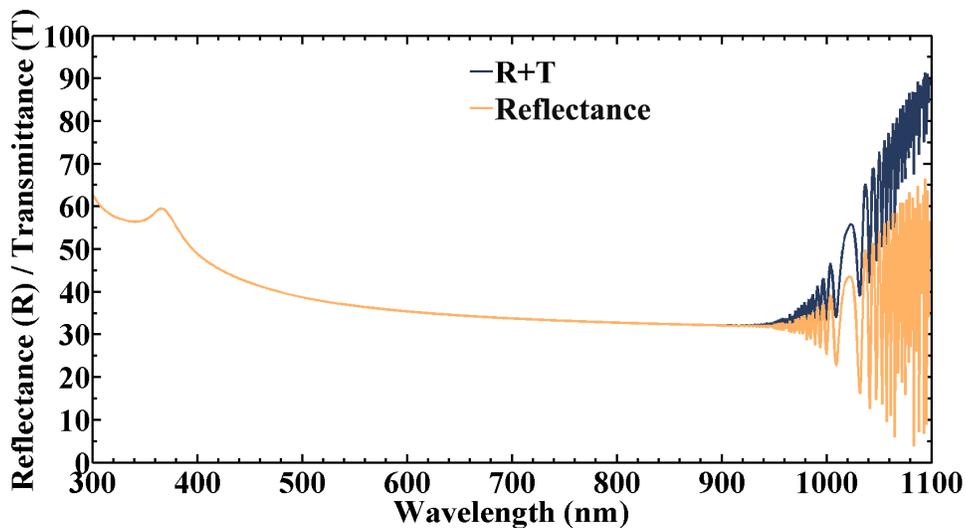

**Figure 2** The spectral reflectance (*R*) and transmittance (*T*) in a 400 μm thick bare silicon wafer.

*2.1 Optical Loss in bare 2 – 400 μm thick c-Si wafers*

The spectral losses for 2 - 400 μm thick planar silicon slabs for the cases in which it has no reflector and it has a hypothetical perfect reflector at the rear surface, are plotted in figure 3(a) and 3(b), respectively. In the latter case we assume no absorption at the back reflector and the spectral losses are entirely due to reflection. With the addition of a the perfect reflector at the back surface, the amplitude of Fabry-Perot peaks increases because of the increase in the reflection from the mirror at the rear surface and the integrated AM1.5g loss only decreases to 62 % compared with 69 % in the case with no back reflector. However, for both cases, the weakly absorbing region extends into the visible region of the spectrum with decreasing silicon thickness. This is indicated by the presence of sharp peaks associated with Fabry-Perot resonance modes in the spectral *R+T* plots. For example, this region extends from 420 nm – 1100 nm in the case of a 2 μm thick silicon foil.

The situation remains the same when we consider absorption of the AM1.5g solar spectrum in the 2 - 400 μm thick planar Si slabs (plotted in figure 3(c)). The optical loss extends into the visible region and increases as the silicon thickness is reduced from 400 μm to 2 μm. The weighted absorption calculated by normalization of the absorption spectrum with the AM1.5g spectrum in the 300 –1107 nm wavelength range decreases from 62.4 % to 37.8 % as the thickness of the Si wafer decreases from 400 μm to 2 μm.

The above observations suggest that the front grating structure should be optimized to facilitate diffraction in the 900-1107 nm range for 400 μm thick Si, and over a broader region, specifically from 450- 1107 nm for 2 μm thick silicon. The front texture should also reduce reflection over a broad wavelength range.

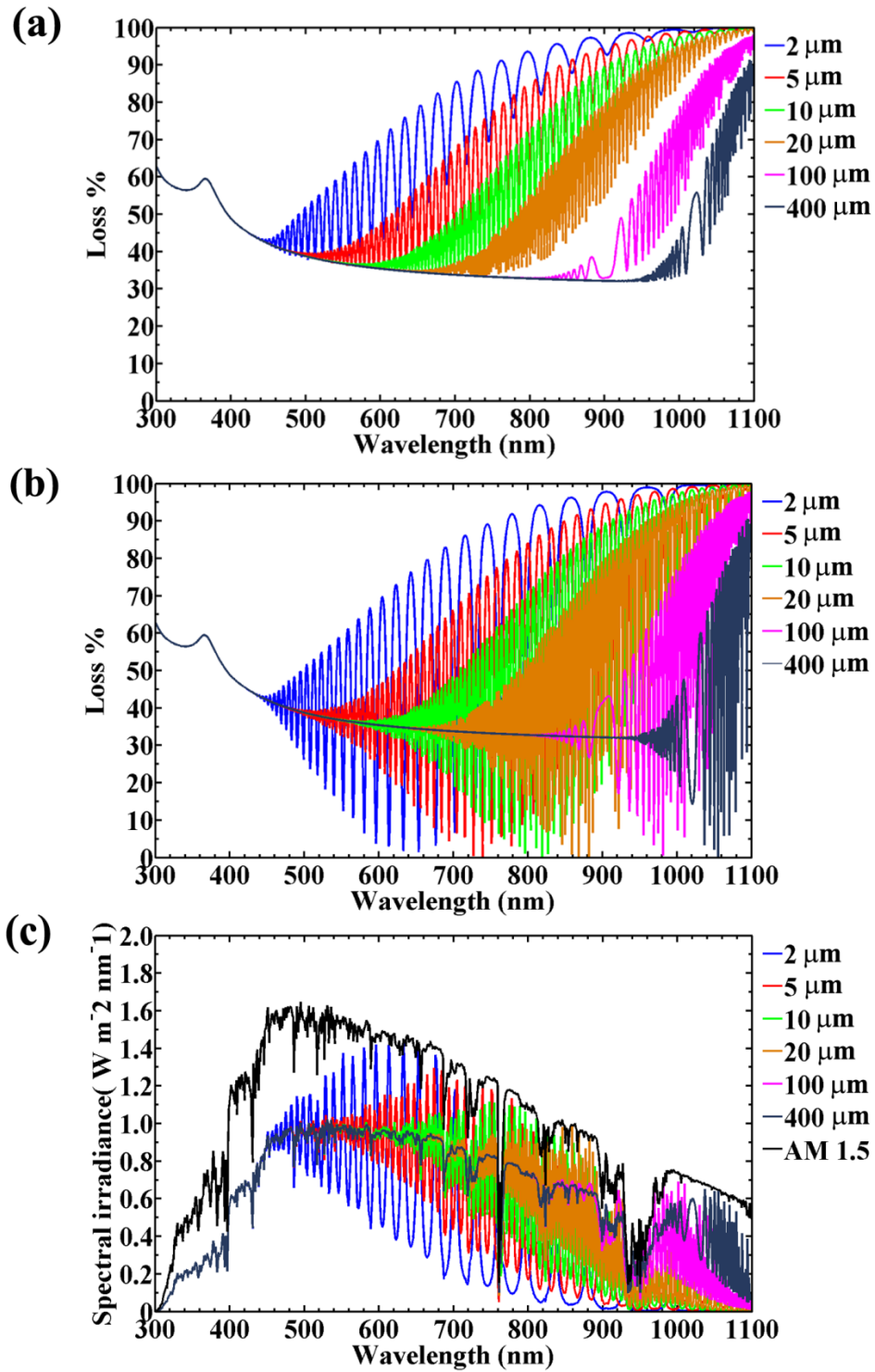

**Figure 3** Total optical loss due to front reflectance and transmittance in a 2 - 400 μm thick silicon (a) without and (b) with a perfect back surface reflector. (c) Absorption of the AM1.5g spectrum in 2- 400 μm thick silicon with a perfect reflector on the rear surface.

3. **Optimal front texture parameters for maximum absorption in 2 – 400 µm thick c-Si wafers**

We now consider adding a tightly packed two-dimensional array of inverted pyramids to the front surface of the slab to enhance photoabsorption. Here we consider the cell architecture shown in Figure 1(c). The calculated photocurrent density in 2 - 400 µm thick silicon with a perfect back reflector and grated front surface with period ($\Lambda$) ranging from 100 – 2000 nm is plotted in figure 4(a). In 400 µm thick silicon, the minimum photocurrent density (31.6 mA/cm$^2$) was calculated for grating period $\Lambda$ = 100 nm. The photocurrent density increased to a local maximum value of 40.8 mA/cm$^2$ as the period is increased to $\Lambda$ = 980 nm, and saturated at 41.2 mA/cm$^2$ for grating periods above $\Lambda$ = 1500. A similar trend in photocurrent was observed for all thicknesses as the grating period was increased from $\Lambda$ = 100 to $\Lambda$ = 1000 nm (figure 4(a)); the photocurrent density increased with increasing grating period and was maximized for $\Lambda \sim$ 1000 nm. The photocurrent density dropped for grating periods larger than 1000 nm except for the case of 400 µm thick silicon where gradual degradation of photocurrent density was observed at very large periods beyond the periodicity scale. The maximization of photocurrent density at $\Lambda \sim$ 1000 nm is expected because such grating periods introduce strong diffraction in the weakly absorbing spectral region just above the absorption edge of Si. When the grating period becomes larger than the wavelengths in the weakly absorbing spectral region, the light couples out of the cell through higher order external modes and decreases the photocurrent density in the cell. This drop in photocurrent density at larger grating periods became more evident in thinner silicon slabs (figure 4(a)) as the relative amount of silicon removed from the available thickness to form the front texture increases significantly as the wafer thickness is reduced from 400 µm to 2 µm. The maximum photocurrent density at a periodicity of $\Lambda \sim$ 1000 nm dropped to 30.79 mA/cm$^2$ with the decrease in thickness of silicon to 2 µm.

A second peak in the photocurrent density occurs at $\Lambda$ = 850 nm for wafer thickness of 10 µm producing 36.45 mA/cm$^2$ photocurrent density similar to 36.46 mA/cm$^2$ at $\Lambda$ = 980 nm. The position of this second peak changed to $\Lambda$ = 650 nm and $\Lambda$ = 680 nm in 5 and 2 µm thick silicon wafers, respectively. However the peak positioned between $\Lambda$ = 980-1000 nm was still present for all thicknesses. The origin of the second peak in the photocurrent density can be derived from the spectral loss curves for different silicon thicknesses shown in figure 3(a)-(c). As previously discussed, with decreasing silicon thickness, optical losses extend to smaller wavelengths in the visible region of the spectrum. In such thin silicon wafers, smaller grating periods in the 650- 850 nm range increase light absorption by reducing surface reflection in the broad spectrum range and facilitate strong diffraction in the visible region of the spectrum. However, we find that the absorption enhancement by smaller grating periods in thin silicon is equivalent to the comparatively large Λ ~ 1000 nm grating period that reduce optical losses mainly by facilitating strong first order and second order diffraction in the IR and visible regions of the spectrum, respectively, and produce similar photocurrent density in silicon.

The peak photocurrent densities at grating periods corresponding to the two peaks in the photocurrent density graph of each silicon thickness with inverted pyramidal front texture (figure 4(a)) and for the ideal case when the grating structure is replaced by a Lambertian surface is plotted in figure 4(b). The maximum photocurrent density generated in 2- 400 μm thick silicon wafers with a grating period of Λ = 1000 nm is close to the photocurrent density generated in wafers of similar thicknesses for the ideal case where the upper surface is Lambertian and a perfect mirror resides at the rear side of the wafer (shown by solid colored lines above the bars). The photocurrent density with grating period corresponding to the peak at (Λ = 680 nm) is only slightly greater than the photocurrent density generated in wafers with Λ = 1000 nm, and this difference is most pronounced for the case of 2 μm thick silicon. The calculated photocurrent density in 10 μm thick silicon wafers with a grating period of 680nm is consistent with the short-circuit current reported by Mavrokefalos et al.[16] for their 10 μm thick silicon samples with a pyramidal front texture with a periodicity of 700nm. From the above results, we conclude that the optimum grating periodicity for an inverted pyramidal front texture to obtain maximum photocurrent density from silicon with an ideal back reflector is Λ = 1000 nm. A high efficiency silicon solar cell can be achieved with this inverted pyramidal grating period, irrespective of silicon wafer thickness.

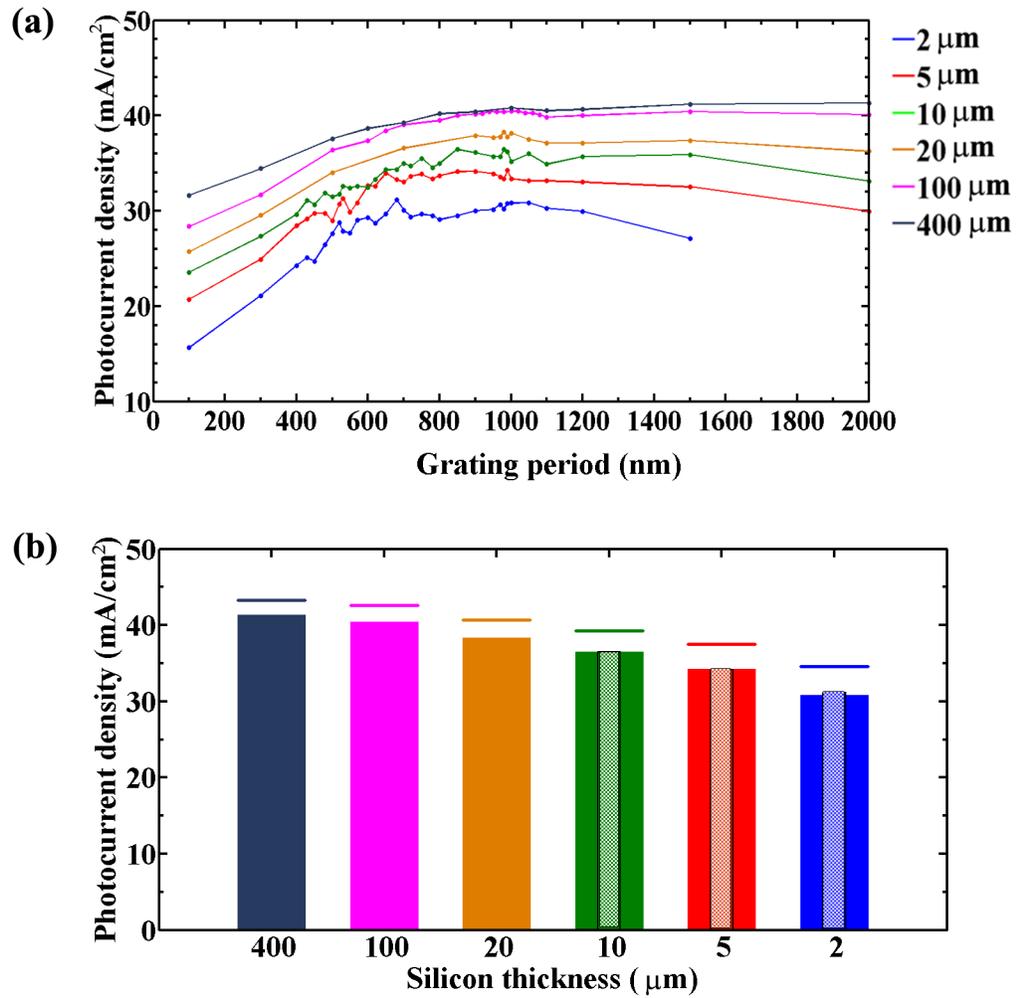

**Figure 4** (a) Simulated photocurrent density in silicon of thickness 2- 400 μm with perfect back reflector and a front inverted pyramidal grating of periodicity ranging from 100 – 2000 nm. (b) Maximum photocurrent density at grating period Λ = 1000 nm (wide solid bars) in 2- 400 μm thick silicon and Λ = 850, 680 and 650 nm in 10, 5 and 2 μm thick silicon, respectively, (narrow faded bars) corresponding to peak positions in (a). Solid lines above the bars represent Yablonovitch limit for each silicon thickness.

4. **Optical loss due to mesas**

Flat ridges (mesas) between the inverted pyramids in the texture as shown in Figure 1(d) are unavoidable irrespective of fabrication process due to the limited accuracy in placing the inverted pyramids precisely with respect to each other. These mesas can be reduced to a practical minimum limit of 100 nm in width in samples prepared with high precision patterning using e-beam lithography. Such flat regions on the surface increase reflection losses and hence decrease the short circuit current in an actual cell. Figure 3(a) shows the photocurrent density in silicon for thicknesses ranging from 2 - 400 μm, respectively, for the optimum grating period Λ ~ 1000 nm with mesa width ranging from 0 - 250 nm. For all thicknesses, the photoabsorption dropped linearly with increasing mesa width at an average rate of 0.02 mA/cm$^2$/nm mesa width.

The optical loss due to mesas can be recovered by adding an anti-reflective coating (ARC) top of the texture. The cell architecture considered here is shown in figure1(e). Our calculations optimize the refractive index of the coating to 2.1 for maximum photoabsorption. This corresponds to materials such as silicon nitride or silicon rich silicon oxide. Figure 5(b) shows the calculated photocurrent density in 2-400 μm thick silicon with optimum front texture parameters with 100 nm mesas coated with different thicknesses of ARC of refractive index 2.1. It is evident that 80 nm thick ARC will maximize the photocurrent density in silicon of all thicknesses.

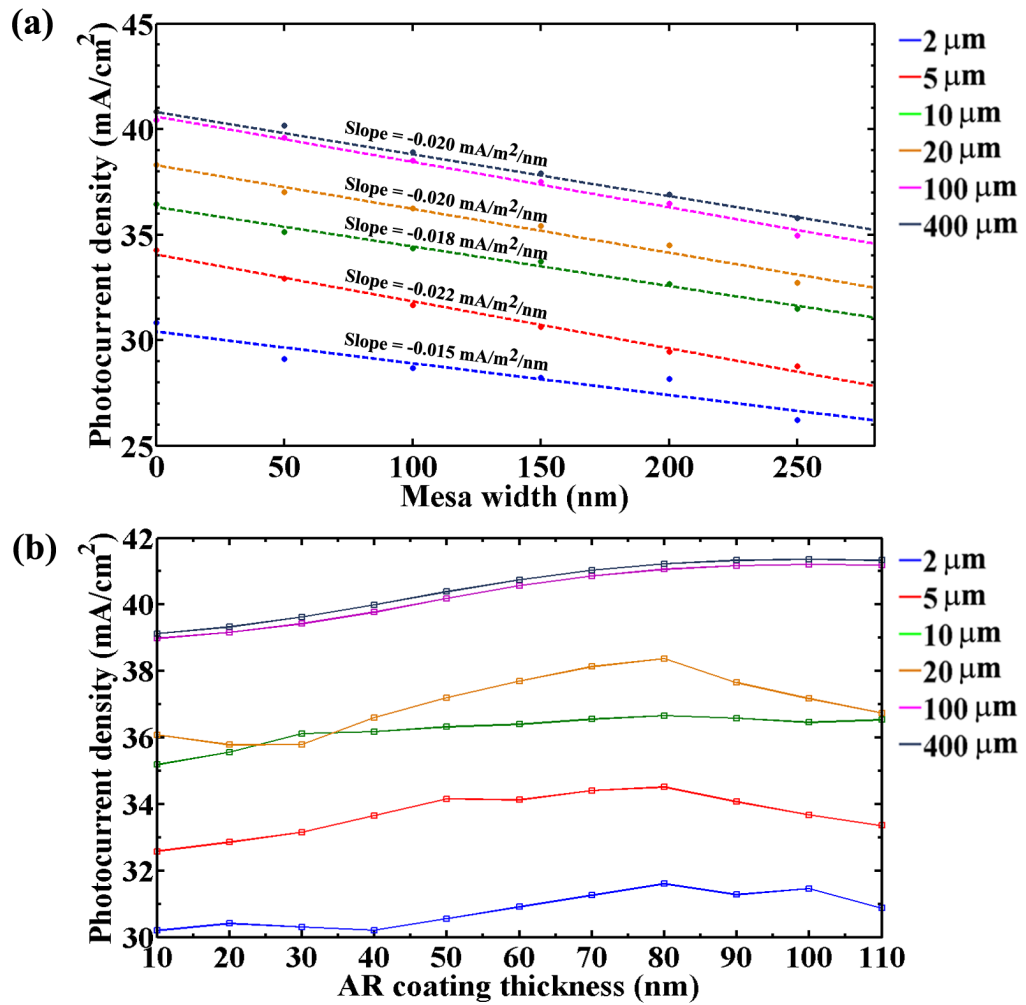

**Figure 3** (a) Degradation in photocurrent density with the increase in mesa width in a grating structure with optimum periodicity of Λ = 1000 nm (b) simulated photocurrent density with optimum texture texture with 100 nm wide mesas coated with different thicknesses of antireflective coating (ARC) of refractive index 2.1 in 2- 400 μm thick crystalline silicon.

Figure 6 shows the photocurrent density in 2 - 400 μm thick silicon for the case when silicon with a perfect reflector at rear surface is textured with an optimal front inverted pyramidal texture with no mesas ( ), with 100 nm wide mesas ( ), 80 nm optimized ARC ( ) and ideal Lambertian texture (×). It is clear that the optimized ARC recovers the optical losses from mesas and also enhances the absorption in the silicon to result in photocurrent densities slightly larger than the case with no mesas for all silicon thicknesses. However, the values were still below the photocurrent density that can be achieved with the ideal Lambertian texture in all cases.

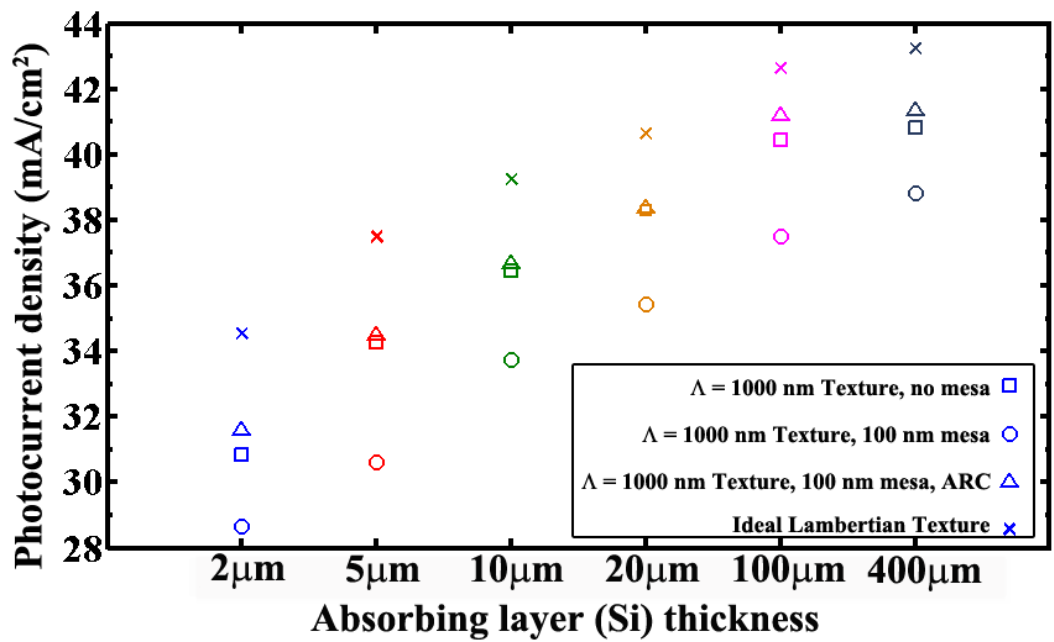

**Figure 6** Simulated photocurrent density current in 2 - 400 μm thick silicon with a perfect reflector at rear surface and optimum texture at the front with no mesas (□) and with 100 nm mesa without (○) and with 80 nm ARC (△), and Lambertian front texture (×).

5. **Experimental Validation**

Electron Beam Lithography (EBL) was used to fabricate 500, 973, and 1500 nm periodic inverted pyramidal textures on 400 μm thick silicon is shown in figure 7. The initial pattern was written in photoresist on a silicon nitride ($SiN_x$) hard mask using an EBPG 5000+ system. The pattern was then transferred to the silicon nitride film by reactive ion etching (RIE). Finally, anisotropic etching was done on the silicon through the holes in the hard mask in order to produce the final inverted pyramidal textured surface with approximately 100 nm wide mesas. The silicon nitride mask was then removed by dipping samples in 1% hydrofluoric acid solution prior to reflectance measurements.

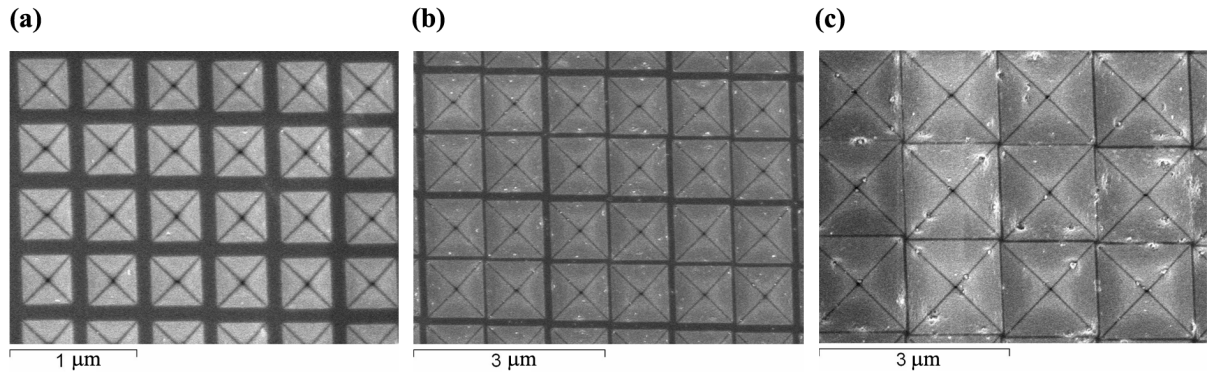

**Figure 7** Scanning electron images of textures fabricated with electron beam lithography with periodicity, $\Lambda$, and feature size, $s$, (a) $\Lambda = 500$ nm, $s = 400$ nm, (b) $\Lambda = 1000$ nm, $s = 900$ nm, (c) $\Lambda = 1500$ nm, $s = 1400$ nm,

The spectral reflectances measured from these three samples with and without a ~ 70 nm thick ARC are shown in figure 8(a) and 8(b), respectively, along with optical simulations at the same texture parameters.

It is clear that the performance of the 500 nm period sample is significantly worse than the larger periodicities in both cases. This is partly because of the large percentage of flat regions in 500 nm sample (fill fraction = 0.8) compared with the 1000 nm (fill fraction = 0.9) and the 1500 nm samples (fill fraction = 0.93). However, this difference alone cannot account for the performance of the 500 nm texture, which performs significantly more than 10% worse than the other periodicities.

The measured spectral reflectance curves of all three textures without an ARC (figure 8(a)) in general shows the same trend with texture size as the measured data, i.e. the 500 nm curve is much higher than the other two. However, the actual level of agreement between data and simulation is not that good. In particular the depth of the troughs (or dips) in the simulated curves are much sharper than in the measurements. This is due to the slight variation in feature size as well as feature alignment in the real samples, since the optical data averages over a very large number of the inverted pyramids. Nevertheless, the dips in the measured reflectance roughly align with the dips predicted by the simulations, as can be seen in the 500 nm curves.

The performance of all three samples is greatly enhanced by an ARC (figure 8(b)), but the 500 nm sample still has the worst performance. The conclusion to be drawn is that from the viewpoint of reflectance there is no particular advantage to making a texture with a period that is as small as 500 nm, i.e. close to the wavelengths of light near the peak of the solar spectrum. The 1000 and 1500 nm period patterns do better, exactly as expected from the curves shown in figure 4. More significantly, the larger periodicity makes EBL unnecessary to fabricate the textured surface in the first place. A 1000 or 1500 nm period texture can be made with optical lithography and deep UV exposure, which would be much more commercially viable.

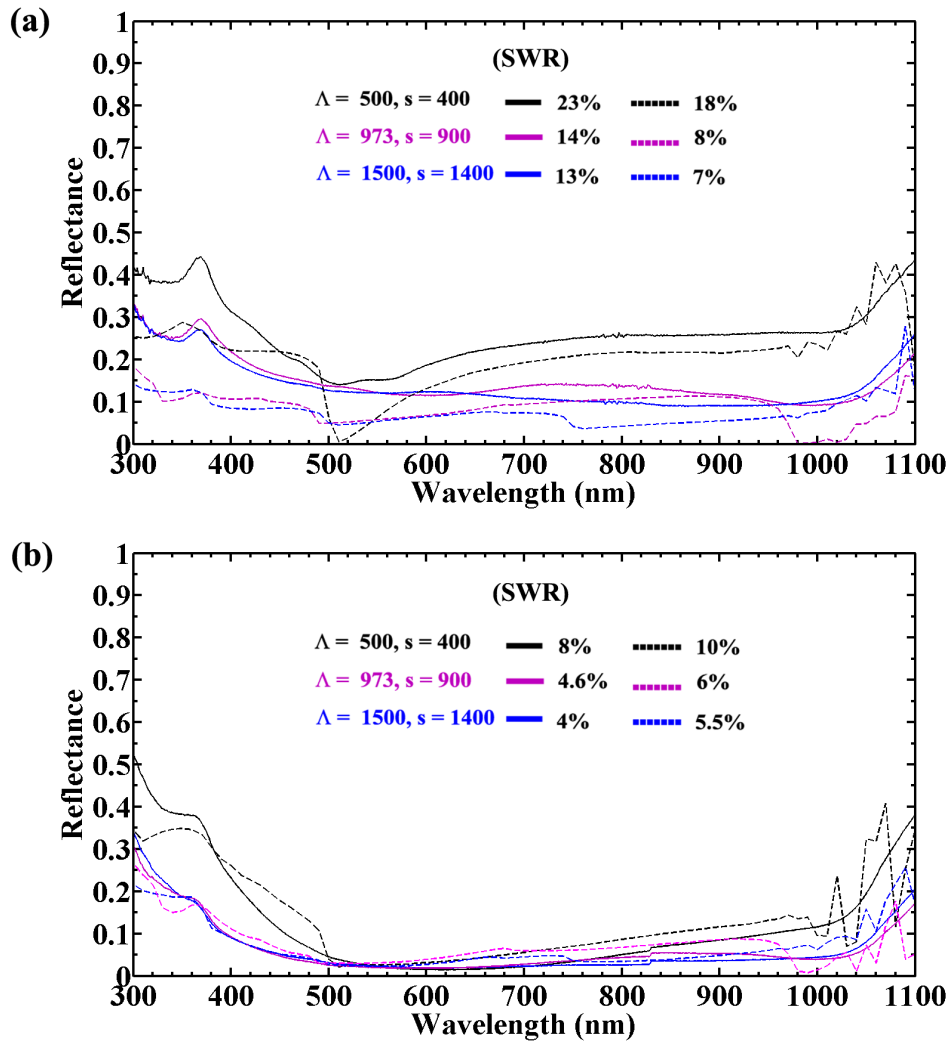

**Figure 8** Simulated (dashed lines) and measured (solid lines) reflectance for three textures with periodicity, $\Lambda$, and feature size, $s$, (a) without and (b) with a 70 nm silicon nitride ARC on a 400 μm thick crystalline silicon wafer. The calculated solar weighted reflectances (SWR) from simulated and experimental reflectance spectra are shown as percentage values.

## 6. Conclusions

The wave-optical study of inverted-pyramidal textures on 2 - 400 μm thick silicon wafers configured with a perfect back reflector shows that a front grating period of 1000 nm leads to maximum photoabsorption with normally incident AM1.5g solar spectrum irrespective of silicon thickness. The calculated photocurrent density with this texture lies close to the Yablonovitch limit for all thicknesses. Our calculations show flat ridges in the texture lead to optical losses and decrease the photocurrent density at the rate of ~ 0.02 mA/cm$^2$/nm mesa width. The reflection loss from mesas can be recovered with the addition of 80 nm thick antireflective coating of 2.1 refractive index.

The reflectance measurements on 500, 973 and 1500 nm periodic inverted pyramidal textures fabricated by electron beam lithography confirms that a 1000 nm scale periodicity results in lower reflectance in comparison with a sub-micron scale periodicity. This result aligns with the theoretical study on reflectance and light trapping study of periodic textures done by Sai *et al*[4], where a periodicity of ≥ 0.8 μm is suggested for textures of features with aspect ratio close to unity. With the identification of such a universally optimized periodicity (Λ = 1000 nm) for the case of an inverted pyramidal grating texture, a common fabrication process can be designed to manufacture high-efficiency devices on crystalline silicon regardless of wafer thickness. Furthermore, the fact that the optimal periodicity is about one micron means that an optimal texturization is also within reach of optical or UV lithography.

# Length Scale Dependence of Periodic Textures for Photoabsorption Enhancement in Ultra-thin Silicon Foils and Thick Wafers

K Kumar[1], A Khalatpour[2], G Liu[1], J Nogami[1] and N P Kherani[1, 2*]

[1] Department of Materials Science and Engineering, University of Toronto, Toronto, M5S 3E4, Canada

[2] Department of Electrical and Computer Engineering, University of Toronto, Toronto, M5S 3G4, Canada

*kherani@ecf.utoronto.ca

## Supplementary Information

Scattering Matrix Method

In this work we use a three-dimensional scattering matrix method [27] to study inverted pyramidal textures with grating periods ranging from subwavelength to ~ 2 × typical wavelengths (300 – 1107 nm) on the front surface of 2 μm to 400 μm thick c-Si with the objective of maximizing photoabsorption of the AM1.5g solar spectrum. Scattering matrix technique has been proposed as an effective tool for solving patterned multilayer structures [28].

In this technique the inverted pyramid grating with periodicity, $\Lambda$, in the lateral (x and y) directions and feature size, $L_z$, (Figure S1) is divided into a set of finite number of layers along the surface normal. The depth, $d$, of the grating in the $z$ direction is defined by the size of the inverted pyramid and is given by $L_z/1.404$. The number of layers is defined by the thickness of each layer, which is set to $\lambda/20$ to ensure accuracy. Each layer is treated as a separate diffraction grating of lateral periodicity $\Lambda$. The field in each layer is ideally expanded in terms of the corresponding infinite set of plane waves with coupled wave vector $k$, which differ by reciprocal lattice vectors $G$, obtained by solving the photonic band structure in each layer. The scattering matrix at each layer is then related to the adjacent layers by enforcing the electromagnetic boundary conditions at the interface of each layer. Consequently, the S-matrix relates the amplitudes of the outgoing waves at the surface and in the substrate, to those of the incoming waves on either side of the structure. The scattering matrix technique is numerically stable compared to the transfer matrix technique, which is widely used for analyzing multilayer unpatterned structures [27]. We encourage interested readers to consult [28] for a detailed explanation of this method.

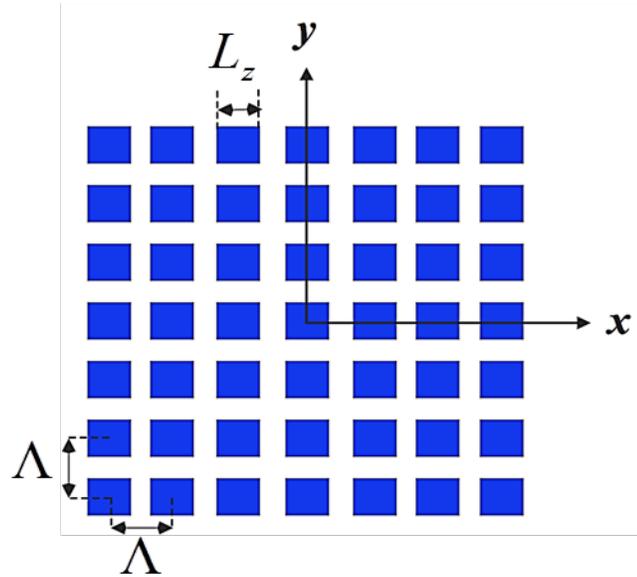

**Figure S1** Top view of inverted pyramidal grating of periodicity, $\Lambda$, in $x$ and $y$ directions and feature size, $L_z$. The depth of the grating in $z$ direction is determined by $L_z$ and is given by $L_z/1.404$.

Ideally, for an accurate calculation all reciprocal vectors, $G$, should be used for wave expansion. However, considering memory and time constraints, in reality numerical calculations avail a set of reciprocal lattice vectors which is truncated at large $G$ values ensuring that the truncation error is within an acceptable range. To ensure accuracy of the solution, the number of lattice vectors in the present calculations was set to 121. Figure S2 shows the calculated reflected power at $\lambda$ = 450 nm (which corresponds to peak power in the AM 1.5g solar spectrum) from a 1000 nm periodic inverted pyramidal texture on a 400 µm thick wafer. The graph shows negligible change for the number of modes equal to or exceeding 81.

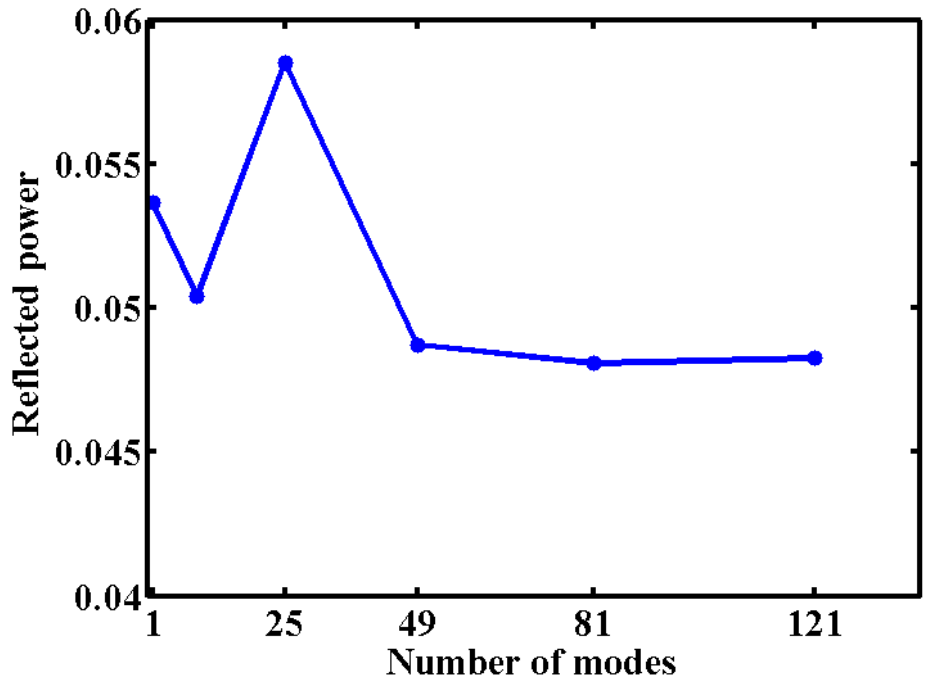

**Figure S2** The calculated reflected power at peak wavelength (λ = 450 nm) in the AM 1.5g solar spectrum when incident on a 1000 nm periodic inverted pyramidal texture on a 400 μm thick wafer. The graph shows negligible change for the number of modes greater than 81.